\documentclass{jfm}
\usepackage{graphicx}
\usepackage{epstopdf}

\usepackage{amsmath}
\usepackage[colorlinks, citecolor=blue]{hyperref}
\usepackage{color}
\usepackage[usenames,dvipsnames]{xcolor} 
\usepackage{setspace}

\shorttitle{Elastocapillary Worthington jets}
\shortauthor{U. Sen, D. Lohse, and M. Jalaal}

\title{Elastocapillary Worthington jets}

\author{Uddalok Sen{\aff{1}$^{,}$\aff{2}}
  \corresp{\email{uddalok.sen@wur.nl}},
  Detlef Lohse{\aff{1}$^{,}$\aff{3}}
  \corresp{\email{d.lohse@utwente.nl}}
   \and Maziyar Jalaal\aff{4}
 \corresp{\email{m.jalaal@uva.nl}}}

\affiliation{\aff{1}Physics of Fluids Group, Max Planck Center Twente for Complex Fluid Dynamics, Department of Science and Technology, and J. M. Burgers Centre for Fluid Dynamics, University of Twente,  P. O. Box 217, 7500 AE Enschede, The Netherlands
\aff{2}Physical Chemistry and Soft Matter, Wageningen University and Research, 6708 WE Wageningen, The Netherlands
\aff{3}Max Planck Institute for Dynamics and Self-Organization, Am Fassberg 17, 37077 G\"{o}ttingen, Germany
\aff{4}van der Waals-Zeeman Institute, University of Amsterdam, Science Park 904, 1098 XH Amsterdam, The Netherlands}

\begin{document}
\maketitle

\begin{abstract}
The retraction of an impacting droplet on a non-wetting substrate is often associated with the formation of a Worthington jet, which is fed by the retracting liquid. A non-Newtonian rheology of the liquid is known to affect the retraction of the impacting droplet. Here we present a novel phenomenon related to the impact of viscoelastic droplets on non-wettable substrates. We reveal that the viscoelasticity of the liquid results in an \emph{elastocapillary} regime in the stretching Worthington jet, distinguished by a pinned contact line and a slender jet that does not detach from the droplet. We identify the impact conditions, in the Weber number -- Deborah number phase space, for observing these \emph{elastocapillary} Worthington jets. Such jets exhibit an effectively nearly linear (in time) variation of the strain rate. Upon further extension, the jet exhibits beads-on-a-string structures, characteristic of the \emph{elastocapillary} thinning of slender viscoelastic liquid filaments. The \emph{elastocapillary} Worthington jet is not only relevant for a droplet impact on a solid substrate scenario, but can also be expected in other configurations where a Worthington jet is observed for viscoelastic liquids, such as drop impact on a liquid pool and bubble bursting at an interface. 
\end{abstract}


\section{Introduction} \label{sec:intro}

The first photographs of a drop impacting on a horizontal substrate published by \citet{worthington-1876-prsa} showed the formation of an eponymous jet, which has fascinated researchers ever since. Such a jet can be formed when a droplet impacts a non-wettable substrate \citep{yarin-2006-arfm, josserand-2016-arfm}. Upon impact, the kinetic energy of the droplet is transformed into surface energy as the droplet expands to attain a maximal spreading diameter, corresponding to a pancake shape. However, on non-wettable substrates, this temporarily-stored surface energy drives the subsequent retraction of the droplet. Capillary waves are excited upon impact, which propagate along the droplet and focus at its axis of symmetry \citep{renardy-2003-jfm, bartolo-2006-prl, zhang-2022-prl}. This results in a flow-focusing effect that culminates in the formation of the so-called \emph{Worthington jet} \citep{zeff-2000-nature}. In essence, these jets formed during drop impact are similar to the jets observed during cavity collapse scenarios \citep{zeff-2000-nature, ghabache-2014-jfm}, such as a bursting bubble near a free interface \citep{duchemin-2002-pof, walls-2015-pre, ghabache-2016-prf, ganancalvo-2017-prl, brasz-2018-prf, deike-2018-prf, lai-2018-prl, gordillo-2019-jfm, ganancalvo-2021-jfm, blancorodriguez-2021-jfm, sanjay-2021-jfm, saade-2021-jfm}, impact on a liquid pool \citep{oguz-1990-jfm, prosperetti-1993-arfm, gekle-2009-prl, ray-2015-jfm,  michon-2017-prf, blancorodriguez-2021-jfm}, an impulsive jet \citep{antkowiak-2007-jfm, kiyama-2016-jfm, gordillo-2020-jfm}, and confined blister collapse \citep{tagawa-2012-prx, peters-2013-jfm, tagawa-2013-labchip, turkoz-2018-prl, jalaal-2019-jfm, quetzerisantiago-2021-softmatter}. These Worthington jets are highly stretched, and often break up via the Rayleigh-Plateau mechanism to form single droplets \citep{bartolo-2006-prl, gordillo-2010-jfm, tagawa-2012-prx, walls-2015-pre, ghabache-2016-prf, ganancalvo-2017-prl, michon-2017-prf, brasz-2018-prf, turkoz-2018-prl}. 

For low viscosity liquids (e.g. water), the post-impact Worthington jet is so violent that the momentum transfer from the drop to the substrate may result in a reaction force that leads to the complete rebound of the droplet  from the substrate \citep{richard-2002-nature, zhang-2022-prl}. This results in loss of liquid retention of the substrate, and is highly detrimental for deposition processes where drop impact is important, such as inkjet printing \citep{lohse-2022-arfm}, spray cooling \citep{kim-2007-intjheatfluidfl}, and pesticide treatment \citep{wirth-1991-pesticsci, hoffman-2021-prf}. Therefore, the recent years have seen the development of several strategies for the rebound suppression of an impacting droplet, such as the addition of surfactants \citep{mourougoucandoni-1997-jcis, hoffman-2021-prf}, the application of an electric field \citep{yun-2013-pre, yun-2014-jfm}, and substrate oscillations \citep{raman-2016-pre, weisensee-2017-prf}. 

One such strategy of rebound suppression that has gained a lot of prominence over the last couple of decades is the addition of polymers to the liquid, thus making the liquids viscoelastic \citep{bergeron-2000-nature, bartolo-2007-prl, smith-2010-prl, chen-2018-macromolecules, dhar-2019-prf, pack-2019-prf}. This is particularly attractive for applications such as spraying pesticides \citep{williams-2008-pestmanagsci, sijs-2020-pestmanagsci, gaillard-2022-jnnfm} and inkjet printing \citep{christanti-2002-jrheol, morrison-2010-rheolacta, yan-2011-pof, sen-2021-jfm}, which in many cases already use viscoelastic working fluids. 

The spreading behavior of an impacting droplet \citep{bartolo-2005-jfm, sen-draft}  is a function of its Weber number, which is the ratio of the inertial to the capillary forces, given by
\begin{equation}
W\!e = \frac{\rho U_{0}^{2} D_{0}}{\gamma}, 
\label{eq:We}
\end{equation}
where $\rho$ is the density of the fluid, $U_{0}$ the impact velocity of a droplet of diameter $D_{0}$, and $\gamma$ the surface tension coefficient, and its Ohnesorge number, which quantifies the ratio of the inertio-capillary to inertio-viscous timescales, 
\begin{equation}
Oh = \frac{\eta}{\sqrt{\rho D_{0} \gamma}}, 
\label{eq:De}
\end{equation}
where $\eta$ is the liquid viscosity. Further, in the context of viscoelastic liquids, one may define the Deborah number as
\begin{equation}
De = \frac{\lambda}{\tau_{\gamma}},
\label{eq:De}
\end{equation}
where $\lambda$ is the relaxation time of the polymers and $\tau_{\gamma} = \sqrt{\rho D_{0}^{3}/8 \gamma}$ is the inertiocapillary timescale. The Deborah number $De$ plays an important role in affecting the pinch-off process of a droplet from a nozzle \citep[see][]{deblais-2020-jfm, sen-2021-jfm}. However, when inertia dominates ($W\!e \gg 1$), the spreading of the impacting droplet is independent of the presence of polymers \citep{bartolo-2007-prl, gorin-2022-langmuir}. In contrast, the retraction process after maximal spreading is affected by the polymers. For Newtonian liquids with $Oh < 0.071$ (and $W\!e \gg 1$), the retraction is driven by capillarity while inertia resists it \citep[inertiocapillary regime, see][]{bartolo-2005-jfm}. The retraction velocity is, in essence, a material constant, and scales with the Taylor-Culick velocity \citep{bartolo-2005-jfm, sanjay-2022-arxiv}, given by
\begin{equation}
v_{TC} = \sqrt{\frac{\gamma}{\rho h_{m}}},
\label{eq:v_TC}
\end{equation}
where $h_{m}$ is the height of the liquid pancake at maximal spreading. For viscoelastic liquids, an additional normal stress at the contact line exists \citep{bartolo-2007-prl, boudaoud-2007-epje}, due to the polymers being stretched by the receding contact line \citep{smith-2010-prl, smith-2014-langmuir, xu-2016-apl, xu-2018-jcis}. This normal stress counters the capillary stress driving the retraction, thus slowing down the retraction process as a whole. This, in turn, suppresses the rebound of the droplets from the substrate. 

Although the motion of the retracting contact line for viscoelastic liquids in drop impact is well-studied \citep{bartolo-2007-prl, smith-2010-prl}, the same cannot be said for the Worthington jet that is formed in those scenarios. The formation and characteristics of a Worthington jet for Newtonian drop impact has been explored in literature \citep{bartolo-2006-prl, chen-2017-langmuir, yamamoto-2018-apl, siddique-2020-prf}, but an understanding of the effect of viscoelasticity on the Worthington jet is still lacking. 

In the present work, we study the drop impact of viscoelastic liquids on non-wettable substrates, with an emphasis on the Worthington jet observed during the retraction of the droplet. This led to the surprising observation of a specific regime where the Worthington jet undergoes a transition from an inertiocapillary regime to a hitherto unknown \emph{elastocapilllary} regime, with the appearance of beads-on-a-string structures. Further, we present a simplified effective model which describes and rationalizes the observed weak time-dependence of the strain rate in that regime. 

The paper is organized as follows: \S~\ref{sec:exp-method} describes the experimental procedure. In \S~\ref{sec:obs}, the phenomenological observations from the experiments are described, culminating in the regime map (figure~\ref{fig:regime map}). In \S~\ref{sec:jet-dyn} we quantify the Worthington jet dynamics from the experiments and extract the temporal variation of the strain rate, which is rationalized in \S~\ref{sec:theory}. The paper ends with conclusions and an outlook in \S~\ref{sec:conclusion}. 

\section{Experimental materials and method} \label{sec:exp-method}

Aqueous solutions of polyethylene oxide (average molecular weight $\approx 10^{6}$ a.u., Sigma-Aldrich, henceforth referred to as PEO), of concentrations $C_{m}$ (by mass) ranging from 0.05\% to 1\% were prepared by adding the required amount of polymer powder to purified water (Milli-Q). For each concentration, a 100 mL solution is made at first, and all experiments were carried out using the same solution. The required polymer amount was measured using a precision laboratory balance (Secura 224-1S, Sartorius) with an accuracy of 0.1 mg, and then added to purified water. Each solution was stirred with a magnetic stirrer for 24 h prior to use in order to ensure homogeneity of concentration. A rotational rheometer (MCR 502, Anton-Paar) with a cone-and-pate configuration (1$^{\circ}$ angle, 50 mm diameter, and mean gap of 0.1 mm) was used to measure the shear viscosities $\eta$ of the solutions (Table~\ref{tab:prop}), as well the polymer ($\eta_{p}$) and solvent ($\eta_{s}$) viscosities. The detailed shear viscosity vs. shear rate curves are shown in figure~\ref{fig:app-rheo}. In the present experiments, all the test liquids were found to be Boger fluids \citep{james-2009-arfm}, i.e. the shear viscosity is independent of the shear rate. The relaxation times $\lambda$ of the PEO solutions were measured from the extensional thinning of liquid filaments in a pendent droplet configuration \citep[see also relaxation time measurement in a Capillary Breakup Extensional Rheometer (CaBER) device: \citet{bazilevskii-1990-erc, yao-1998-jnnfm, amarouchene-2001-prl, anna-2001-jrheol, clasen-2006-jfm}]{deblais-2018-prl, mathues-2018-jor, sur-2018-jor, deblais-2020-jfm}. The surface tension coefficient $\gamma$ was measured by the pendent droplet method in an optical contact angle measurement and contour analysis instrument (OCA 15, Dataphysics). From Table~\ref{tab:prop}, it can be noted that the surface tension coefficients $\gamma$ and densities $\rho$ remain unchanged with concentration (except for pure water, $C_{m}$~=~0), while changes with concentration were observed in the measurements of viscosity $\eta$ and relaxation time $\lambda$. The overlap concentration $C_{m}^{\ast}$ for PEO1M was calculated to be 0.135\% \citep{clasen-2006-jor, tirtaatmadja-2006-pof}, and the corresponding $C_{m}/C^{\ast}_{m}$ values for each of the concentrations used in the present experiments have been mentioned in Table~\ref{tab:prop}.

\begin{table}
\begin{center}
\def~{\hphantom{0}}
\begin{tabular}{lllllll}
$C_{m}$ [wt. \%] & $\eta$ [mPa.s] & $\gamma$ [N/m] & $\lambda$ [ms] & $\rho$ [kg/m$^{3}$] & $\eta_{p}/\eta_{s}$ & $C_{m}/C_{m}^{\ast}$ \\[3pt]
0 (pure water) & 0.89 & 0.072 & 0 & 1000 & 0 & N/A \\
0.05 & 0.91 & 0.060 & 0.137 & 1000 & 0.013 & 0.37 \\
0.10 & 1.15 & 0.060 & 0.192 & 1000 & 0.281 & 0.74 \\
0.25 & 1.49 & 0.060 & 0.447 & 1000 & 0.659 & 1.85 \\
0.50 & 3.02 & 0.060 & 0.917 & 1000 & 2.363 & 3.70 \\
0.75 & 8.24 & 0.060 & 1.952 & 1000 & 8.176 & 5.56 \\
1.00 & 14.83 & 0.060 & 2.547 & 1000 & 15.514 & 7.41
\end{tabular}
\caption{Salient properties of the PEO solutions used in the present work.}
\label{tab:prop}
\end{center}
\end{table}

For preparing the superhydrophobic substrates, microscope glass slides (76 mm $\times$ 26 mm, Menzel Gl\"{a}ser) were first sonicated in acetone (Boom) in an ultrasound bath sonicator (USC600T, VWR) for 15 mins, then rinsed (with purified water) and dried, and subsequently treated in an oxygen plasma cleaner (Harrick Plasma Inc.) for 3 mins. A coating of silanized silica nanobeads of 20 nm diameter (Glaco MirrorCoat Zero) was then applied to render the glass slide superhydrophobic. A dip coating procedure was used, where the glass slide was dipped vertically in the coating suspension and then slowly pulled out. The coated slide was then baked in a convection oven at 120$^{\circ}$C oven for 1 h. This dipping-and-baking process was repeated two more times to obtain a superhydrophobic glass slide on which a water droplet displayed an apparent static contact angle of 160$^{\circ}$, measured with an optical contact angle and contour analysis instrument (OCA 15, Dataphysics). In the present work, we focus only on superhydrophobic substrates in the present work for two reasons. Firstly, the work is motivated by applications in inkjet printing and pesticide treatment of plants -- applications where the substrate might often be non-wettable (e.g. flex advertising banners and lotus leaves). Secondly, in \S~\ref{sec:jet-dyn} and \S~\ref{sec:theory}, we focus on the dynamics of the Worthington jet, the formation of which requires the droplet to recoil after the post-impact expansion stage -- which only happens if the substrate is non-wettable.

\begin{figure}
\centering
\includegraphics[width=\textwidth]{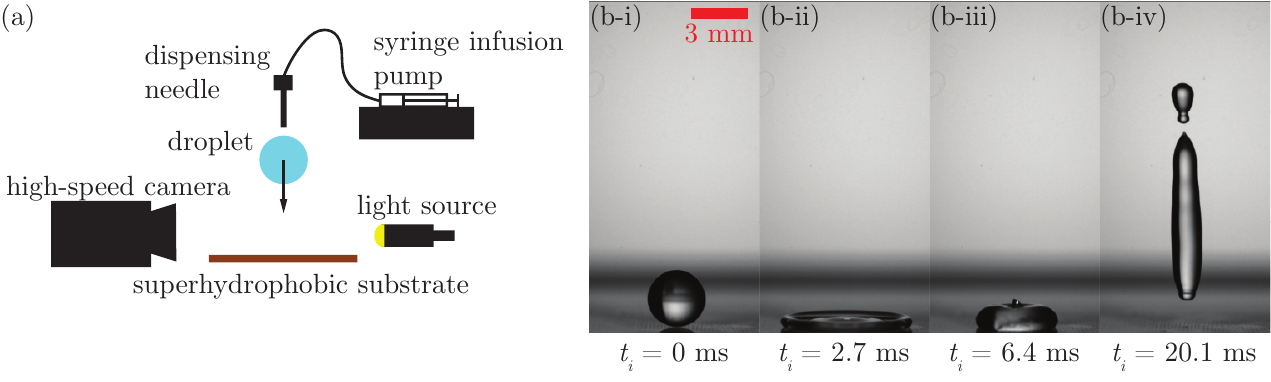}
\caption{(a) Schematic of the experimental setup; (b-i) -- (b-iv) typical time-lapsed snapshots of a drop impacting a superhydrophobic substrate ($W\!e = 54.88$, $De= 0.023$, $Oh$ = 0.002; \emph{rebound} regime). See movie SM2 in the supplementary material for the corresponding movie.}
\label{fig:setup}
\end{figure}

A schematic of the experimental setup is shown in figure~\ref{fig:setup}a. A teflon-coated stainless steel needle (inner diameter = 0.25 mm, Nordson EFD) was used to dispense droplets of diameter 2.6 mm. The droplets were generated by connecting the needle to a syringe (Injekt 5 mL, Braun) via a flexible plastic PEEK tubing (Upchurch Scientific). The syringe was actuated by operating a syringe pump (Harvard Apparatus) in the infusion mode at a very low flow rate (0.1 mL/min). This ensured that the droplet pinched off from the needle with negligible velocity. The height of the needle from the substrate was varied in order to change the impact velocity of the droplet. The impact was visualized from the side by a high-speed camera (at $10^{4}$ frames per second, Nova S12, Photron) connected to a macro lens (50 mm DG macro prime, Sigma) with 56 mm of lens extension tube (Kenko). This resulted in an optical resolution of 17.24 $\mu$m/pixel. The impact events were back-illuminated by a cold LED light source (at 100\% intensity, KL 2500 LED, Schott) through a plane diffuser sheet. A typical drop impact event is shown in figure~\ref{fig:setup}b (further elaborated upon in \S~\ref{sec:obs}). The obtained high-speed images were then analyzed using an OpenCV-based Python script developed in-house. The script utilized a thresholding technique to binarize the raw grayscale images. The binarized images were further analyzed to obtain the salient dimensions during the drop impact event. For all the datasets shown below, the error bars ($\pm$ one standard deviation, calculated from five independent experimental realizations) are smaller than the symbols representing the datapoints. Hence, they are not shown for clarity. Apart from $W\!e$, $Oh$, and $De$ defined in \S~\ref{sec:intro}, the other dimensionless group of importance in the present problem is the elastocapillary number $Ec$, defined as
\begin{equation}
Ec = \frac{G D_{0}}{\gamma},
\label{eq:Ec}
\end{equation}
which represents the ratio of the elastic to capillary stresses, and where $G$ denotes the elastic modulus.

\section{Phenomenological observations} \label{sec:obs}

\begin{figure}
\centering
\includegraphics[width=\textwidth]{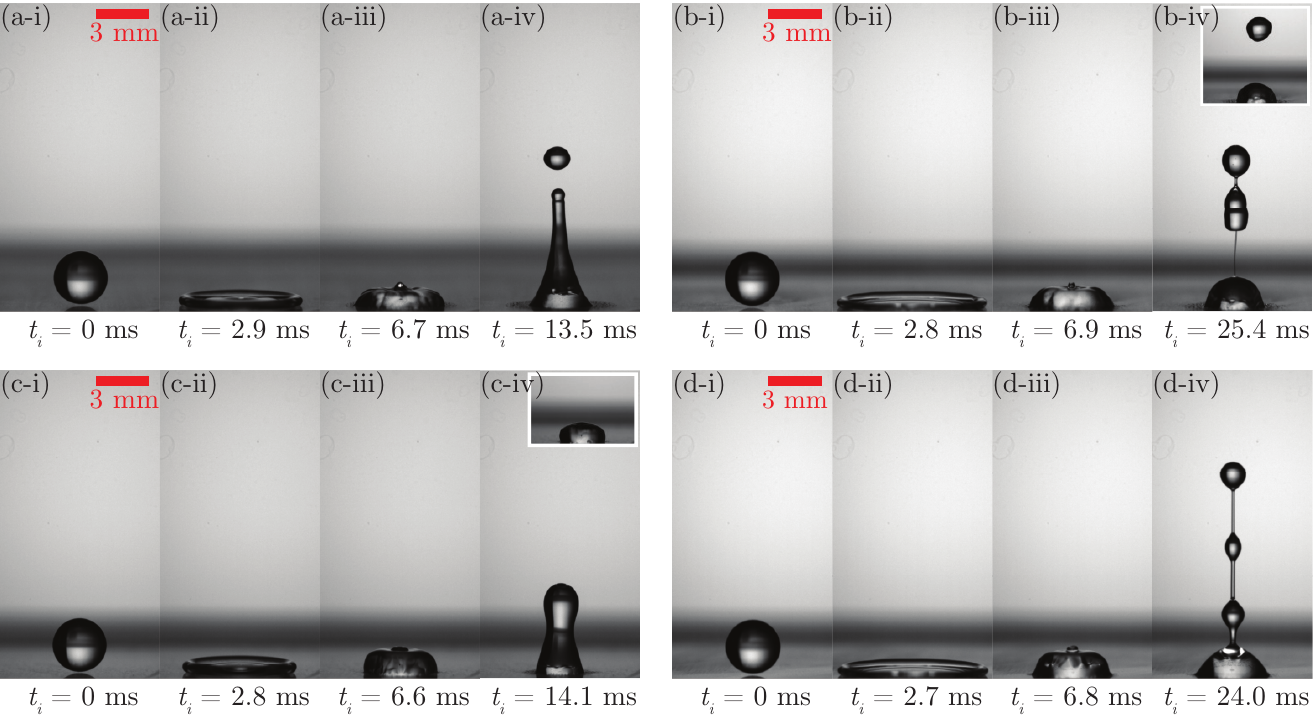}
\caption{Typical time-lapsed snapshots of the drop impact event on a superhydrophobic substrate for the (a) \emph{partial rebound} regime ($W\!e = 54.88$, $De = 0.075$, $Oh = 0.003$), (b) \emph{partial rebound + beads} regime ($W\!e = 104.09$, $De = 0.158$, $Oh = 0.004$), (c) \emph{deposition} regime ($W\!e = 54.88$, $De = 0.431$, $Oh = 0.038$), and (d) \emph{deposition + beads} regime ($W\!e = 112.25$, $De = 0.326$, $Oh = 0.021$). See movies SM1 and SM2 in the supplementary material for the corresponding movies.}
\label{fig:observations}
\end{figure}

\begin{figure}
\centering
\includegraphics[width=\textwidth]{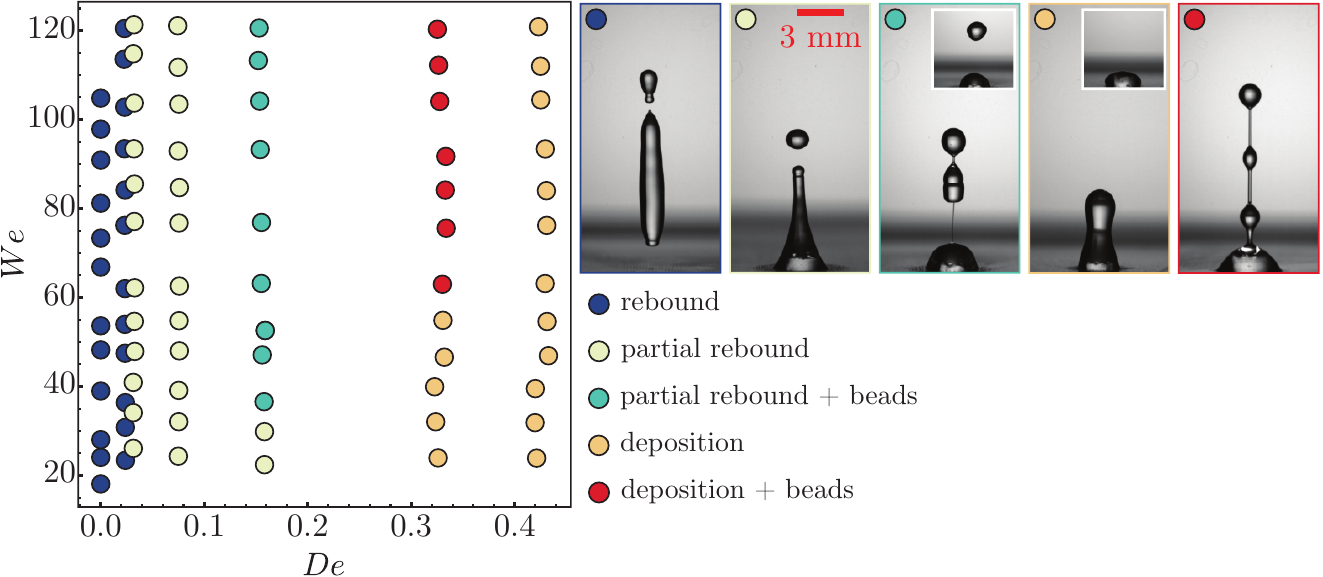}
\caption{Regime map based on our experiments in the $W\!e$ -- $De$ phase space for the drop impact of viscoelastic liquids, along with the typical snapshots of each regime. See movies SM1 and SM2 of the supplementary material for the corresponding movies. Identifying the exact crossovers between the various regimes is beyond the scope of the present article.}
\label{fig:regime map}
\end{figure}

In this work, the PEO concentration $C_{m}$ and the drop impact velocity $U_{0}$ are independently varied. This results in the observation of five different regimes in the $W\!e$ -- $De$ phase space (figure~\ref{fig:regime map}): \emph{rebound} (figure~\ref{fig:setup}b), \emph{partial rebound} (figure~\ref{fig:observations}a), \emph{partial rebound + beads}~(figure \ref{fig:observations}b), \emph{deposition}~(figure \ref{fig:observations}c), and \emph{deposition + beads}~(figure \ref{fig:observations}d). For all the regimes, the droplet impacts and spreads to a maximal diameter. This expansion phase of the droplet \citep{bartolo-2005-jfm} is independent of $De$ for a fixed $W\!e$ \citep[see figure~\ref{fig:app-spreading} and][]{bartolo-2007-prl, gorin-2022-langmuir}. The liquid pancake formed at maximal spreading subsequently retracts, and it is this retraction phase that is affected by varying $De$ \citep[see~figure \ref{fig:app-spreading} and][]{bartolo-2005-jfm}, leading to the different regimes in figures~\ref{fig:observations} and \ref{fig:regime map}. 

In the \emph{rebound} regime (blue datapoints in figure~\ref{fig:regime map}), the droplet completely detaches from the substrate (figure~\ref{fig:setup}b and movie SM2 of the supplementary material), and is reminiscent of water droplets impacting a superhydrophobic substrate \citep{yarin-2006-arfm, josserand-2016-arfm}. Upon increasing $De$, the \emph{partial rebound} regime (yellow datapoints in figure~\ref{fig:regime map}) is observed, where a part of the droplet leaves the substrate while the rest of the droplet remains pinned (figure~\ref{fig:observations}a and movie SM2 of the supplementary material). From the \emph{partial rebound} regime onwards, pinning of the receding contact line is observed during later times (see figure \ref{fig:app-spreading}, where the spreading diameter $D/D_{0}$ remains almost constant at late times), also reported in earlier studies of viscoelastic drop impact \citep{bergeron-2000-nature, bartolo-2007-prl, smith-2010-prl, yang-2022-softmatter}. The normal force at the contact line hindering the retraction increases with increasing polymer concentration \citep{bartolo-2007-prl}. Hence, the pinned contact line appears in our experiments when $De$ is increased. In the present experiments, the \emph{rebound} regime was observed for a highest $De$ = 0.023, while the first instance of the \emph{partial rebound} regime was observed at $De$ = 0.031. That being said, it must be noted that these values do not indicate the precise location of the transition between these two regimes. The identification of the transitions between the different regimes is not within the scope of the present work.

One observes the first signature of beads-on-a-string structures in the part of the droplet detaching from the substrate as $W\!e$ increases in the \emph{partial rebound} regime ($De = 0.158$ and $W\!e \gtrsim 40$ in figure~\ref{fig:regime map}, figure~\ref{fig:observations}b, and supplementary movie SM2). This regime, where we observe beads-on-a-string structures on the detaching part of the droplet, is henceforth denoted as the \emph{partial rebound + beads} regime. Beads-on-a-string morphologies are commonly observed in an intermediate thinning regime of extensional flows of viscoelastic liquids \citep{goldin-1969-jfm, bousfield-1986-jnnfm, wagner-2005-prl, clasen-2006-jfm, oliveira-2006-jnnfm, deblais-2018-prl}, termed \emph{elastocapillary} regime \citep{bazilevskii-1990-erc, entov-1997-jnnfm, anna-2001-jrheol, amarouchene-2001-prl, deblais-2018-prl, eggers-2020-jfm, deblais-2020-jfm}. Such structures are a direct consequence of the viscoelasticity-induced extended lifetime of a slender viscoelastic filament undergoing capillary thinning. The radius of the filament continuously decreases, and fluid is forced by capillarity from the thin regions of the filament into the spherical beads \citep{clasen-2006-jfm}. 

On increasing $De$ further, rebound is completely suppressed, and the \emph{deposition} regime is observed (orange datapoints in figure~\ref{fig:regime map}). In this regime, a Worthington jet is formed but droplets do not detach from the jet (figure~\ref{fig:observations}c and supplementary movie SM2). Within this regime where droplet rebound is suppressed, one observes that at $De~=~0.326$ and $W\!e~\gtrsim~60$, the Worthington jet forms beads-on-a-string structures without any droplets pinching off (figure~\ref{fig:observations}d, and supplementary movies SM1 and SM2). We call this the \emph{deposition + beads} regime (red datapoints in figure~\ref{fig:regime map}).

The primary difference between the \emph{partial rebound} (or \emph{partial rebound + beads}) and \emph{deposition} (or \emph{deposition + beads}) regime is that the Worthington jet does not break up in the latter case. The Worthington jet in our experiments is essentially a viscoelastic filament under extension \citep{gekle-2010-jfm, gordillo-2010-jfm}. For such a filament, the thinning of the filament radius as it is stretched is initially identical to Newtonian fluids. After sufficient extension, the dissolved polymers undergo a coil-to-stretch transition \citep{eggers-2020-jfm}. Once the finite extensibility limit of the polymers is reached, the filament again thins like a Newtonian fluid till it ruptures \citep{entov-1997-jnnfm, fontelos-2004-jnnfm, bhat-2008-jnnfm}. Note that the experimental determination of this finite extensibility limit is a challenge, and usually requires several (unknown) fitting parameters \citep{lindner-2003-physicaa, sen-2021-jfm}. Hence, we focus our discussion to the \emph{deposition} and \emph{deposition + beads} regimes, where the finite extensibility limit of the polymers has not been reached (i.e. the jet does not break up). Further, it is known that the time of onset of finite extensibility depends linearly on the relaxation time $\lambda$, which explains the delayed onset of finite extensibility at higher $De$, resulting in the \emph{deposition} and \emph{deposition + beads} regimes. We devote the rest of this paper on studying the Worthington jet dynamics, in order to identify what distinguishes the \emph{deposition + beads} regime from the \emph{deposition} regime. 

\section{Worthington jet dynamics} \label{sec:jet-dyn}

\begin{figure}
\centering
\includegraphics[width=\textwidth]{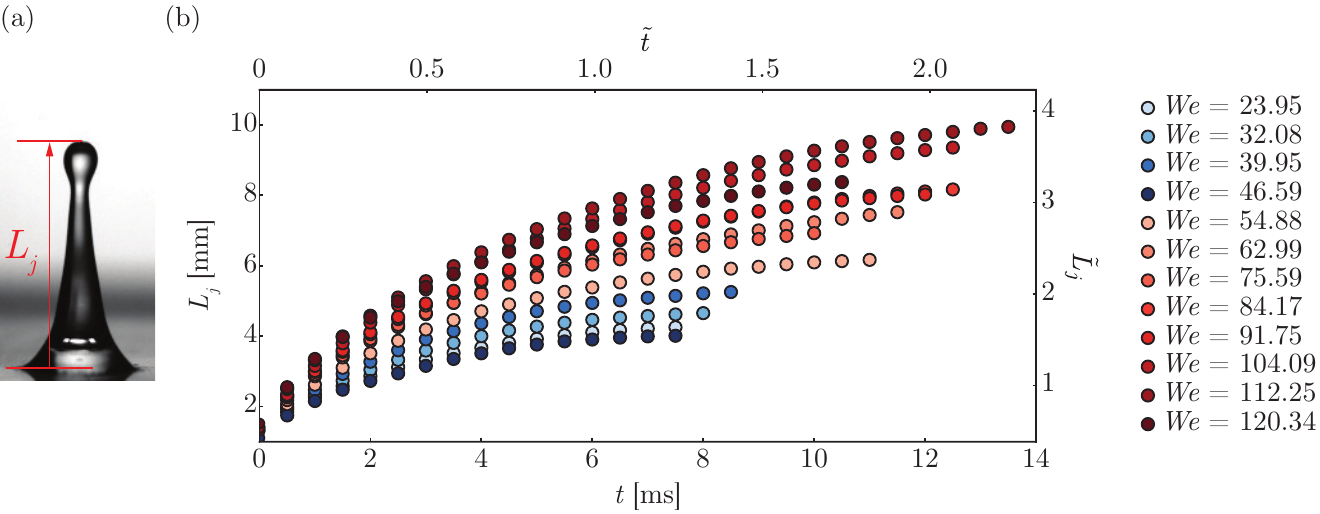}
\caption{(a) A typical measurement of the jet length $L_{j}$ from the experimental snapshots. (b) Variation of the jet length with time for different $W\!e$ in the dimensional $L_{j}$($t$) (left and lower axes) and non-dimensional $\tilde{L}_{j}$($\tilde{t}$) (right and upper axes) forms, where $t$ = 0 ms (or $\tilde{t}$ = 0) indicates the time where the jet first protrudes above the liquid pancake. The blue color gradient corresponds to the \emph{deposition} regime while the red color gradient denotes the \emph{deposition + beads} regime (see figures~\ref{fig:observations} and \ref{fig:regime map}). For all the datapoints shown here: $De = 0.326$ and $Oh = 0.021$.}
\label{fig:exp dynamics}
\end{figure}

For all our experiments, $Oh < 0.071$. Therefore, the droplet retraction that leads to the formation of the Worthington jet is inertiocapillary in nature \citep{bartolo-2005-jfm}. In particular, we analyze the jet dynamics for $De = 0.326$. The Worthington jet is then driven by inertia while capillarity and elasticity resist it. We infer the jet dynamics from the temporal evolution of the jet length $L_{j}$. A typical measurement of $L_{j}$ from the experimental snapshots is shown in figure~\ref{fig:exp dynamics}a, while the temporal variation of $L_{j}$ for $De = 0.326$ and varying $W\!e$ is shown in figure~\ref{fig:exp dynamics}b (left and lower axes). Here, $t = 0$ denotes the time instant when the jet first protrudes above the liquid pancake (figure~\ref{fig:observations}d-iii). In the discussion that follows, all lengths are normalized by the droplet diameter $D_{0}$, and all times are normalized by the inertiocapillary time $\tau_{\gamma} = \sqrt{\rho D_{0}^{3}/8 \gamma}$. The variation of the normalized jet length $\tilde{L}_{j} = L_{j}/D_{0}$ with normalized time $\tilde{t} = t/\tau_{\gamma}$ is also shown in figure~\ref{fig:exp dynamics}b (right and upper axes). Each dataset in figure~\ref{fig:exp dynamics}b is truncated to times before the maxima of $L_{j}$ (and/or $\tilde{L}_{j}$) are reached. Furthermore, in figure~\ref{fig:exp dynamics}, the datasets in blue color gradient correspond to the \emph{deposition} regime while those in the red color gradient denote the \emph{depostion + beads} regime. It can be observed from figure~\ref{fig:exp dynamics}b that the jet length $\tilde{L}_{j}$ increases in a sub-linear manner with time $\tilde{t}$. Moreover, jets in the \emph{deposition + beads} regime (red color gradient) extend longer than the jets in the \emph{deposition} regime. 

We will analyze the stretched Worthington jet as a slender viscoelastic jet under uniaxial tension. For such uniaxial stretching, the normalized strain rate can be expressed as 
\begin{equation}
\dot{\tilde{\varepsilon}} \sim \frac{\partial \tilde{u}}{\partial \tilde{z}} \sim \frac{1}{\dot{\tilde{L}}_{j}} \frac{\mathrm{d} \dot{\tilde{L}}_{j}}{\mathrm{d} \tilde{t}} \sim \frac{\ddot{\tilde{L}}_{j}}{\dot{\tilde{L}}_{j}} ,
\label{eq:eps-sim}
\end{equation}
where $\tilde{z}$ and $\tilde{u}$ are the axial coordinate and velocity, respectively, and a dot denotes differentiation with respect to normalized time $\tilde{t}$. In (\ref{eq:eps-sim}), we have chosen $\tilde{L}_{j}$ to be the scale for $\tilde{z}$ and $\dot{\tilde{L}}_{j}$ to be the scale for $\tilde{u}$. Since $L_{j}$ can be directly measured from our experiments, this allows us to get an estimate of $\dot{\tilde{\varepsilon}}$. Further, $\tilde{L}_{j}$ only depends on $\tilde{t}$ and not on $\tilde{z}$ in the present experiments. It is to be noted that (\ref{eq:eps-sim}) is consistent with the limit of Newtonian Worthington jets exhibiting ballistic behavior \citep[where the momentum equation can be written as the material derivative of axial velocity field to be zero; valid when the tip droplet has not formed, see][]{gekle-2010-jfm}. Further, upon invoking a one-dimensional mass balance along the axis of the jet, (\ref{eq:eps-sim}) can also be written as 
\begin{equation}
\dot{\tilde{\varepsilon}} \sim \frac{\ddot{\tilde{L}}_{j}}{\dot{\tilde{L}}_{j}} \sim \frac{1}{\dot{\tilde{L}}_{j}} \frac{\mathrm{d} \dot{\tilde{L}}_{j}}{\mathrm{d} \tilde{t}} \sim \frac{\mathrm{d} \ln \dot{\tilde{L}}_{j}}{\mathrm{d} \tilde{t}} \sim 2 \frac{\mathrm{d} \ln \tilde{r}}{\mathrm{d} \tilde{t}} \sim \frac{\dot{\tilde{r}}}{\tilde{r}},
\label{eq:eps-r}
\end{equation}
where $r(t)$ is the radius of the jet.

\begin{figure}
\centering
\includegraphics[width=\textwidth]{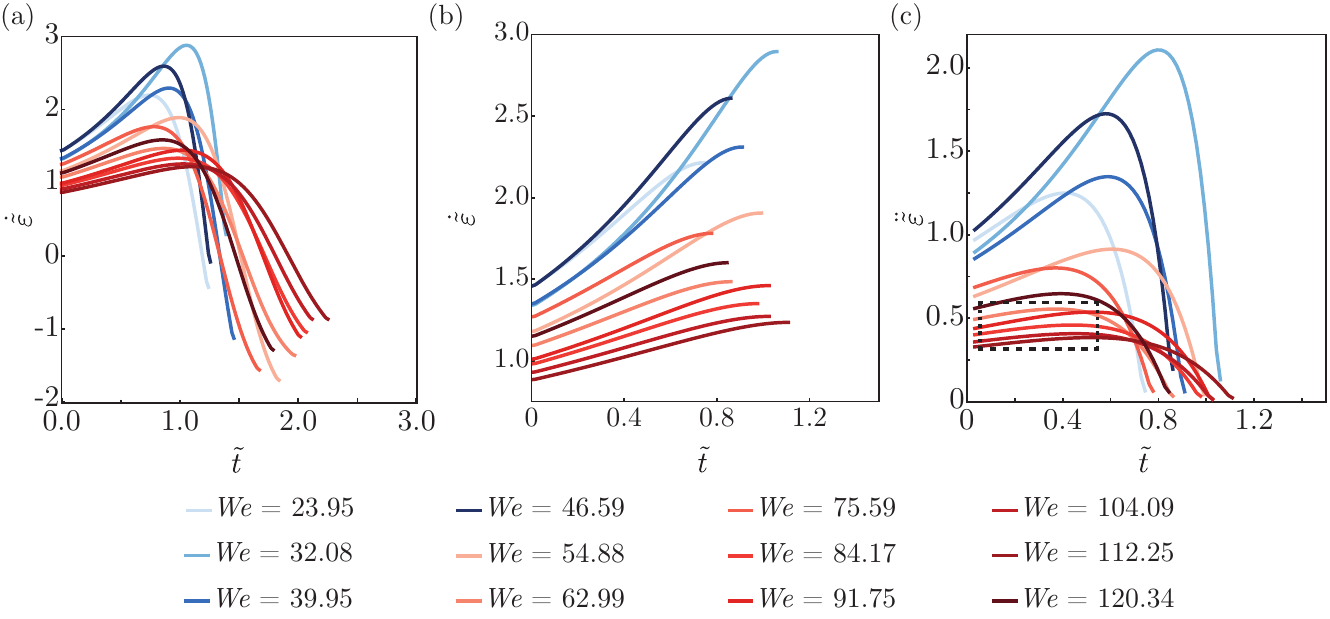}
\caption{(a) Strain rate $\dot{\tilde{\varepsilon}}$, calculated from the dataset in figure~\ref{fig:exp dynamics}b, plotted as function of $\tilde{t}$ for different $W\!e$. (b) The same dataset as in panel (a) but truncated at the maximum value of each curve for better focus on the initial behavior. (c) $\ddot{\tilde{\varepsilon}} = \mathrm{d} \dot{\tilde{\varepsilon}} / \mathrm{d} \tilde{t}$, calculated from the curves in panel (b), plotted as a function of $\tilde{t}$ for different $W\!e$. The black dashed rectangle in panel (c) highlights the weak $\tilde{t}$-dependence of $\ddot{\tilde{\varepsilon}}$ at early times for large $W\!e$. The blue color gradient corresponds to the \emph{deposition} regime while the red color gradient denotes the \emph{deposition + beads} regime (see figures~\ref{fig:observations} and \ref{fig:regime map}). For all the datapoints shown here: $De = 0.326$ and $Oh = 0.021$.}
\label{fig:calc dynamics}
\end{figure}

For the datasets shown in figure~\ref{fig:exp dynamics}b, $\dot{\tilde{\varepsilon}} = - \ddot{\tilde{L}}_{j} / \dot{\tilde{L}}_{j}$ is calculated by fitting an univariate spline through the $\dot{\tilde{L}}_{j} (\tilde{t})$ data. The temporal variation of $\dot{\tilde{\varepsilon}}$ (figure~\ref{fig:calc dynamics}a) shows an initial increase followed by a steep decrease for all the curves. The steepness is more pronounced for the \emph{deposition} regime (blue color gradient) than for the \emph{deposition + beads} regime (red color gradient). This sharp decrease is due to the formation of the top droplet (and intermediate beads in the \emph{deposition + beads} regime), where (\ref{eq:eps-sim}) is not valid. Hence, we will focus on the $\dot{\tilde{\varepsilon}}(\tilde{t})$ variation only up to the maxima of each curve (figure~\ref{fig:calc dynamics}b), where a monotonous increase with $\tilde{t}$ is observed. The \emph{deposition} curves (blue color gradient) and \emph{deposition + beads} curves (red color gradient) show different characteristics of $\dot{\tilde{\varepsilon}}$ at early times: the former appears to have a much stronger dependence on $\tilde{t}$ than the latter. To investigate this further, we plot $\ddot{\tilde{\varepsilon}} = \mathrm{d} \dot{\tilde{\varepsilon}}/\mathrm{d} \tilde{t}$ against $\tilde{t}$ in figure~\ref{fig:calc dynamics}c. It can be observed from figure~\ref{fig:calc dynamics}c that $\ddot{\tilde{\varepsilon}}$ is largely time-independent for small times (black dashed rectangle in figure~\ref{fig:calc dynamics}c) for the \emph{deposition + beads} regime (red color gradient), while there is still a strong time-dependence of $\ddot{\tilde{\varepsilon}}$ at early times for the \emph{deposition} (blue color gradient) regime. This further indicates a stronger time-dependence of $\dot{\tilde{\varepsilon}}$ for the \emph{deposition} regime than for the \emph{deposition + beads} regime, readily observed in figure~\ref{fig:calc dynamics}b. 

\section{Simple effective model} \label{sec:theory}

\begin{figure}
\centering
\includegraphics[width=\textwidth]{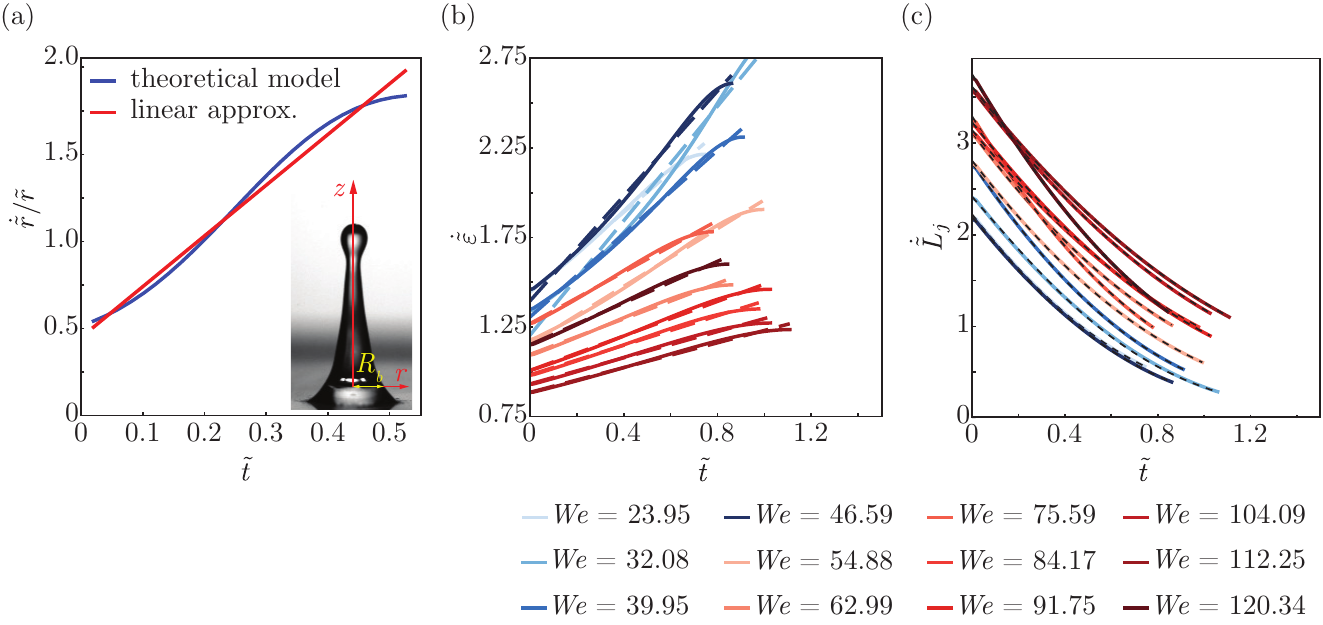}
\caption{(a) $\dot{\tilde{\varepsilon}} \sim \dot{\tilde{r}}/\tilde{r}$ as a function of $\tilde{t}$ as obtained from the numerical solution of~(\ref{eq:r-evol}): the blue curve shows the numerical solution of the theoretical model while the red line is a linear approximation; inset shows the axisymmetric coordinate system for the slender jet analysis in \S~\ref{sec:theory}. (b) Linear fits (dashed lines) to the dataset shown in figure~\ref{fig:calc dynamics}b. (c) Variation of $\dot{\tilde{L}}_{j}$ as function of $\tilde{t}$, where the solid lines represent the experimental dataset while the dashed lines denote the fits~(\ref{eq:vel-fit}). For panel (b) and (c): the blue color gradient corresponds to the \emph{deposition} regime while the red color gradient denotes the \emph{deposition + beads} regime (see figures \ref{fig:observations} and \ref{fig:regime map}). For all the datapoints in panels (b) and (c): $De = 0.326$ and $Oh = 0.021$.}
\label{fig:theory}
\end{figure}

We further analyze the stretching Worthington jet in an axisymmetric coordinate system (inset of figure~\ref{fig:theory}a). The slender jet formulation is adopted \citep{eggers-1993-prl, eggers-1994-jfm, shi-1994-science}, which assumes that the length of the jet is much larger than its radius, and the axial velocity dominates over the radial velocity. We hypothesize that the jet is in the \emph{elastocapillary} regime, where only capillarity and viscoelasticity govern the dynamics of the jet. For any cross-section along the jet, the net tensile force is given by \citep{clasen-2009-jfm}
\begin{equation}
\tilde{F} = F/\gamma D_{0} = \pi \tilde{r} + \pi \tilde{r}^{2} \tilde{\sigma}_{p} ,
\label{eq:force}
\end{equation}
where $\tilde{\sigma}_{p} = (\sigma_{p} D_{0} / \gamma)$ is the (non-dimensionalized) polymeric stress. We assume that at the base of the jet (radius $\tilde{R}_{b}$), the polymers are unstretched or fully-relaxed \citep{clasen-2009-jfm}. Hence, there is no polymeric stress contribution to the force at the base of the jet, and the net tensile force at that location is given by
\begin{equation}
\tilde{F} = \pi \tilde{R}_{b} .
\label{eq:force-base}
\end{equation}
We further assume that the tensile force along the jet is a constant \citep{clasen-2009-jfm}. It has been shown by \citet{clasen-2006-jfm, bhat-2012-pof, eggers-2020-jfm} using the slender jet approximation and the Oldroyd-B model for polymer stresses that in the elastocapillary limit, the total tensile force inside a thinning filament is independent of the axial coordinate $\tilde{z}$, and only depends on the time $\tilde{t}$. This allows us to equate (\ref{eq:force}) and (\ref{eq:force-base}), leading to
\begin{equation}
\tilde{\sigma}_{p} = \frac{1}{\tilde{r}} \left( \frac{\tilde{R}_{b}}{\tilde{r}} - 1 \right) .
\label{eq:stress-r}
\end{equation}
For the \emph{deposition + beads} regime in our experiments, $De = 0.326$ and $Oh = 0.021$. Thus we neglect additional viscous stresses in our analysis, while elastic stresses are considered since they become significant at large strains. 

For a slender cylindrical filament \citep{entov-1997-jnnfm}, 
\begin{equation}
\dot{\tilde{\varepsilon}}_{zz} = - 2 \, \frac{\dot{\tilde{r}}}{\tilde{r}}, \quad \dot{\tilde{\varepsilon}}_{rr} = \frac{\dot{\tilde{r}}}{\tilde{r}}
\label{eq:eps-slenderjet}
\end{equation}
are the axial and radial components of the strain rate tensor $\underline{\underline{\dot{\tilde{\varepsilon}}}}$, respectively. We choose the FENE-CR model \citep{book-bird} to represent the polymeric stresses. This model has been successfully used before to study the thinning of slender viscoelastic filaments \citep{entov-1997-jnnfm, clasen-2009-jfm}. In the FENE-CR model, the components of the conformation tensor $\underline{\underline{A}}$ are expressed as
\begin{equation}
\dot{A}_{zz} = 2 \, \dot{\tilde{\varepsilon}}_{zz} \, A_{zz} - \frac{f}{De} \left( A_{zz} - 1 \right) ,
\label{eq:Azz-fene}
\end{equation}
\begin{equation}
\dot{A}_{rr} = 2 \, \dot{\tilde{\varepsilon}}_{rr} \, A_{rr} - \frac{f}{De} \left( A_{rr} - 1 \right) ,
\label{eq:Arr-fene}
\end{equation}
where $f$ is the correction term accounting for the finite extensibility of the polymer molecules, given by
\begin{equation}
f = \left( 1 + \frac{3}{b} - \frac{A_{zz} + 2 \, A_{rr}}{b} \right)^{-1} ,
\label{eq:f-fene}
\end{equation}
$b$ being the limit of $A_{zz}$ at the maximum extension of the polymer molecules. The components of the polymeric stress tensor can then be expressed as 
\begin{equation}
\tilde{\sigma}_{zz} = 4 \sqrt{2} \, Oh \, \dot{\tilde{\varepsilon}}_{zz} + Ec \, f \, \left( A_{zz} - 1 \right) ,
\label{eq:sigma-zz-fene}
\end{equation}
\begin{equation}
\tilde{\sigma}_{rr} = 4 \sqrt{2} \, Oh \, \dot{\tilde{\varepsilon}}_{rr} + Ec \, f \, \left( A_{rr} - 1 \right). 
\label{eq:sigma-rr-fene}
\end{equation}

The polymeric stress $\tilde{\sigma}_{p}$ in (\ref{eq:force}) is then given by
\begin{equation}
\tilde{\sigma}_{p} = \tilde{\sigma}_{zz} - \tilde{\sigma}_{rr} = - 12 \sqrt{2} \, Oh \, \frac{\dot{\tilde{r}}}{\tilde{r}} + Ec \, f\left( A_{zz} - A_{rr} \right) .
\label{eq:sigma-p-fene}
\end{equation}
Equating the expressions of $\tilde{\sigma}_{p}$ from (\ref{eq:stress-r}) and (\ref{eq:sigma-p-fene}), we get
\begin{equation}
12 \sqrt{2} \, Oh \, \frac{\dot{\tilde{r}}}{\tilde{r}} = - \frac{\tilde{R}_{b}}{\tilde{r}^{2}} + \frac{1}{\tilde{r}} + Ec \, f \left( A_
{zz} - A_{rr} \right).
\label{eq:r-evol-fene}
\end{equation}

In this analysis, we focus on slender jets under strong axial stretching, where the polymers do not reach their finite extensibility limit. In such a scenario:
\begin{equation}
A_{zz} \gg 1, \quad A_{zz} \gg A_{rr}, \quad A_{zz} \ll b, \quad f \approx 1.
\label{eq:approx}
\end{equation}
Therefore, (\ref{eq:r-evol-fene}) and (\ref{eq:Azz-fene}) can be simplified to
\begin{equation}
12 \sqrt{2} \, Oh \, \frac{\dot{\tilde{r}}}{\tilde{r}} = - \frac{\tilde{R}_{b}}{\tilde{r}^{2}} + \frac{1}{\tilde{r}} + Ec \, A_{zz} ,
\label{eq:r-evol-approx}
\end{equation}
\begin{equation}
\dot{A}_{zz} + 4 \, \frac{\dot{\tilde{r}}}{\tilde{r}} \, A_{zz} = - \frac{1}{De} A_{zz}.
\label{eq:Azz-approx}
\end{equation}
Since the polymers are assumed to be unstretched at the base of the jet, we can further write \citep{clasen-2009-jfm}
\begin{equation}
A_{zz} \, \tilde{r}^{4} = A_{zz} \vert_{\tilde{z} = 0} \, \tilde{R}_{b}^{4} ,
\label{eq:Azz-base}
\end{equation}
where $A_{zz} \vert_{(\tilde{z} = 0, \tilde{t} = 0)} = 1$ by definition. The relation between the conformation of the polymers at any $\tilde{z}$ location to that at the base of the jet is expressed by \eqref{eq:Azz-base}, where $A_{zz} \sim \tilde{L}_{j}^{2}$ \citep{sen-2021-jfm}, and one can then use one-dimensional mass balance to relate $A_{zz}$ and $\tilde{r}$. Therefore, using (\ref{eq:Azz-base}) to solve (\ref{eq:Azz-approx}) leads to an exponential decay of $A_{zz}$ given by
\begin{equation}
A_{zz} \, \tilde{r}^{4} = \tilde{R}_{b}^{4} \, e^{- \tilde{t}/De}.
\label{eq:Azz-exp}
\end{equation}
Substituting (\ref{eq:Azz-exp}) into (\ref{eq:r-evol-approx}) reveals an approximate governing equation for the temporal variation of the jet radius:
\begin{equation}
12 \sqrt{2} \, Oh \, \frac{\dot{\tilde{r}}}{\tilde{r}} = - \frac{\tilde{R}_{b}}{\tilde{r}^{2}} + \frac{1}{\tilde{r}} + Ec \, \frac{\tilde{R}_{b}^{4}}{\tilde{r}^{4}} \, e^{-\tilde{t}/De} .
\label{eq:r-evol}
\end{equation}
This equation can now be solved numerically with the initial condition $\tilde{r} \vert_{\tilde{t} = 0} = \tilde{R}_{b}$ to obtain the nonlinear temporal variation of $\dot{\tilde{\varepsilon}} \sim \dot{\tilde{r}}/\tilde{r}$ (\ref{eq:eps-r}), as shown by the blue curve in figure~\ref{fig:theory}a. 

Interestingly, this temporal variation of $\dot{\tilde{\varepsilon}}$ can be reasonably approximated to be effectively linear (red line in figure~\ref{fig:theory}a) -- which is similar to what we observe from our experiments ($\ddot{\tilde{\varepsilon}}$ largely $\tilde{t}$-independent at small times, figure~\ref{fig:calc dynamics}c). This is also what one expects from the elastocapillary thinning of an infinitely long slender cylindrical filament \citep{entov-1997-jnnfm}, where the thinning of the jet is exponential in time, thus indicating that the strain rate is linear in time. Therefore, we conclude that the Worthington jet observed in the \emph{deposition + beads} regime (figure~\ref{fig:observations}d) are, in fact, the hitherto unknown \emph{elastocapillary} Worthington jets, whose governing equation \eqref{eq:r-evol} describes a balance between elasticity and capillarity that governs the jet dynamics. The governing equation \eqref{eq:r-evol} also indicates that the thinning of an elastocapillary Worthington jet follows an $\exp(-\tilde{t}/De)$ decay, whereas the infintitely long slender cylindrical jet studied by \citet{entov-1997-jnnfm} (and several others over the years) thins at a rate of $\exp(-\tilde{t}/3 De)$. We note that the plateauing of the $\dot{\tilde{\varepsilon}}(\tilde{t})$ curves observed at later times in figure \ref{fig:theory}b is due to the pronounced axial non-uniformity of the jet radius, where the present analysis is no longer valid. 

It is to be noted that the droplet diameter is chosen as the length scale in the present experiments for calculating $De$ since that can be controlled easily in the experiments. However, for the Worthington jet that is formed post-impact, the relevant length scale rather is the jet diameter, which is smaller than the droplet diameter in the present experiments. Although the current values of $De$ based on the droplet diameter are all $<$ 1 (which may apparently mean that elastic effects are not as relevant as capillary effects), the value of $De$ based on jet diameter would be $\gtrsim 1$, thus signifying that elastic effects are as relevant as capillary effects. However, calculating the exact value of $De$ based on the jet diameter is difficult in the present experiments since the jet diameter is not axially-uniform. Further, in contrast to the droplet diameter, the jet diameter is not a control parameter of the experiments, and depends on the various impact parameters of the droplet. Hence, we chose to define $De$ based on the droplet diameter. 

At this point, we would like to obtain an effective functional form of $\tilde{L}_{j}(\tilde{t})$, which is difficult since we do not have an analytical expression of $\dot{\tilde{\varepsilon}}(\tilde{t})$. However, from the above analysis, we have shown that $\dot{\tilde{\varepsilon}}$ varies approximately linearly with $\tilde{t}$ in the \emph{elastocapillary} regime. We can thus fit
\begin{equation}
\dot{\tilde{\varepsilon}} = A + B \, \tilde{t}
\label{eq:epsfit}
\end{equation}
to our experimentally-obtained $\dot{\tilde{\varepsilon}}(\tilde{t})$ curves (figure~\ref{fig:calc dynamics}b). The fits are shown in figure~\ref{fig:theory}b, while the variation of the fitting coefficients $A$ and $B$ with $W\!e$ are shown in figures~\ref{fig:app-fitting}a and \ref{fig:app-fitting}b, respectively. It can be identified from figure~\ref{fig:theory}b that the linear fits (dashed lines) match reasonably well with the experimental data (solid lines) in the \emph{deposition + beads} regime (red color gradient), while the fits are poor in the \emph{deposition} regime (blue color gradient). 

In order to account for the $W\!e$-dependence, we can now solve
\begin{equation}
\frac{1}{\dot{\tilde{L}}_{j}} \frac{\mathrm{d} \dot{\tilde{L}}_{j}}{\mathrm{d} \tilde{t}} = - \dot{\tilde{\varepsilon}} = - A - B \, \tilde{t}
\label{eq:vel-diffeq}
\end{equation}
to get the approximate temporal variation of $\dot{\tilde{L}}_{j}$, namely
\begin{equation}
\dot{\tilde{L}}_{j} = K \exp \left( - A \, \tilde{t} - \frac{B}{2} \tilde{t}^{2} \right) ,
\label{eq:vel-fit}
\end{equation}
where $K$ denotes the initial jet velocity (at $\tilde{t} = 0$). The effective description (\ref{eq:vel-fit}) is fitted to the experimental datapoints, see figure~\ref{fig:theory}c.  

For completeness, we show in figure \ref{fig:app-fitting} that the fitting parameters $A$, $B$, and $K$ depend on the Weber number $W\!e$. The variation of $K$ with $W\!e$ is consistent with a $W\!e^{1/3}$ scaling (figure~\ref{fig:app-fitting}c), but given the limited range of our data, we refrain from claiming any scaling relationship at this point. We only mention this because the same dependence $K \sim W\!e^{1/3}$ of the initial jet velocity on $W\!e$ was reported recently in a study of Newtonian drop impact \citep{zhang-2022-prl}. However, a mechanistic understanding of the dependence of the initial jet velocity on the impact $W\!e$ is still lacking in the scientific literature. 

\section{Conclusions and outlook} \label{sec:conclusion}

In this paper, we studied the post-impact behavior of viscoelastic droplets on superhydrophobic substrates. Our experiments reveal several unique regimes based on the behavior of the Worthington jet in the $W\!e$ -- $De$ phase space. A particularly fascinating scenario is the \emph{deposition + beads} regime, which is characterized by a pinned contact line of the droplet, and a Worthington jet that stretches to form beads-on-a-string structures. We show that such a Worthington jet can be approximated to have an effectively linear (in time) dependence of the strain rate, which can be rationalized from a simple model based on the slender jet formulation. Crucially, the simple model does not explicitly contain any inertia term; it describes a balance of elasticity and capillarity, unlike the traditional Worthington jet known in literature, which is a balance of inertia and capillarity \citep[see][]{gekle-2010-jfm}. We propose that this is an \emph{elastocapillary} Worthington jet, unlike the traditional inertiocapillary Worthington jet. To the best of our knowledge, \emph{elastocapillary} Worthington jets have not been reported before. Once in the \emph{elastocapillary} regime, nonlinear effects have to be taken into account to study the formation of the beads-on-a-string structures on stretching viscoelastic filaments \citep{clasen-2006-jfm, bhat-2010-natphys}, which goes beyond our simple effective model. Analysis of the formation of beads-on-a-string structures requires detailed numerical investigations \citep{clasen-2006-jfm, bhat-2010-natphys, ardekani-2010-jfm}, which is beyond the scope of the present work. 

Our findings are not only relevant for industrial viscoelastic droplet deposition processes such as they occur in inkjet printing and in the treatment of plants with pesticides, but are also of interest in other configurations where Worthington jets are produced with viscoelastic liquids, such as bubble bursting at a free interface, impact on a liquid pool, impulsive jets, and confined blister collapse. 

\noindent{\textbf{Acknowledgments}}

The authors thank Andrea Prosperetti, Vatsal Sanjay, and Jacco Snoeijer for fruitful discussions. 

\noindent{\textbf{Funding}}
 
We acknowledge the funding by an Industrial Partnership Programme of the Netherlands Organisation for Scientific Research (NWO), co-financed by Canon Production Printing B. V., University of Twente, and Eindhoven University of Technology. \\

\noindent{\textbf{Declaration of interests}}

The authors report no conflict of interest. \\

\noindent{\textbf{Supplementary information}} \label{SM}

Supplementary information is available at (URL to be inserted by publisher). \\

\noindent{\textbf{Author ORCID}}

U. Sen \href{https://orcid.org/0000-0001-6355-7605}{https://orcid.org/0000-0001-6355-7605};  

D. Lohse \href{https://orcid.org/0000-0003-4138-2255}{https://orcid.org/0000-0003-4138-2255};

M. Jalaal \href{https://orcid.org/0000-0002-5654-8505}{https://orcid.org/0000-0002-5654-8505}

\appendix

\section{Post-impact expansion and retraction} \label{sec:app-spreading}

\begin{figure}
\centering
\includegraphics[width=\textwidth]{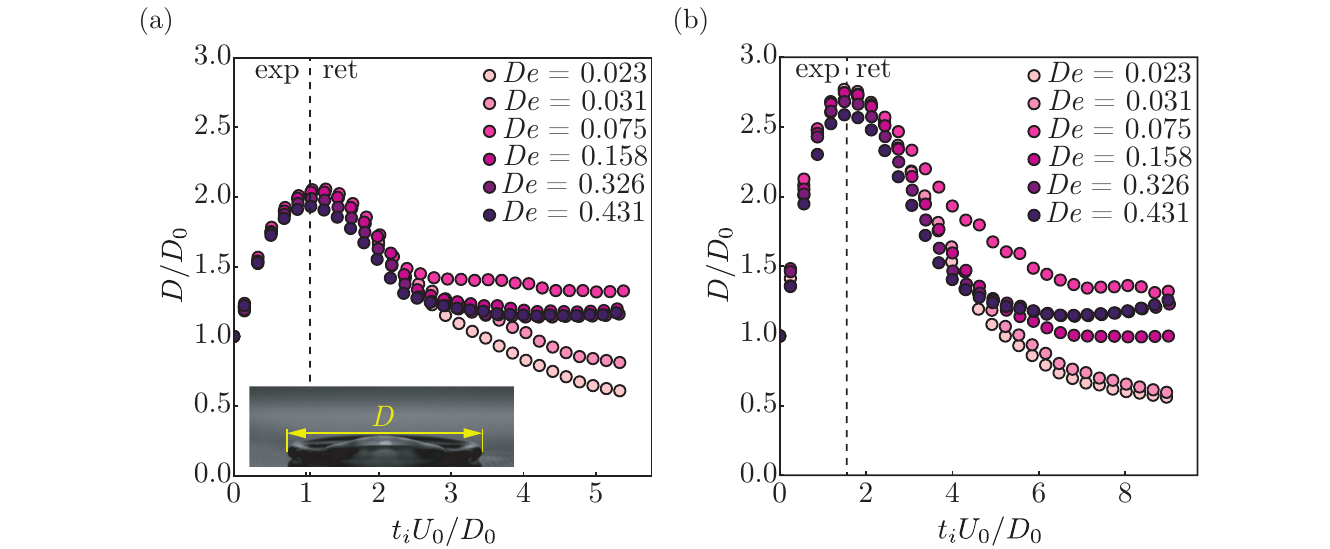}
\caption{Variation of the normalized spreading diameter (=$D/D_{0}$) with normalized time ($= t_{i} U_{0} / D_{0}$) for different $De$ at (a) $W\!e = 39.95$ and (b) $W\!e = 112.25$, showing the expansion (exp) and retraction (ret) phases. The inset of panel (a) shows a typical measurement of $D$ from the experimental snapshots.}
\label{fig:app-spreading}
\end{figure}

The evolution of the spreading diameter $D$ of the droplet (inset of figure~\ref{fig:app-spreading}), normalized by the droplet diameter $D_{0}$, is plotted against time $t_{i}$, normalized as $t_{i} U_{0}/D_{0}$ , in figure~\ref{fig:app-spreading} for two different $W\!e$ (= 39.95 and 112.25) and six different $De$ (= 0.023, 0.031, 0.075, 0.158, 0.326, and 0.431). The plots have two phases: the expansion phase (`exp', from impact to maximal spreading diameter) and the retraction phase (`ret', beyond the maximal spreading diameter). The expansion phase is observed to be independent of $De$, and only depends on $W\!e$ in the present experiments. The retraction phase, on the other hand, is strongly affected by $De$. Such observations were also reported in previous studies of viscoelastic drop impact \citep{bartolo-2007-prl, gorin-2022-langmuir}. 

\section{Initial velocity of the Worthington jet} \label{sec:app-fitting}

\begin{figure}
\centering
\includegraphics[width=\textwidth]{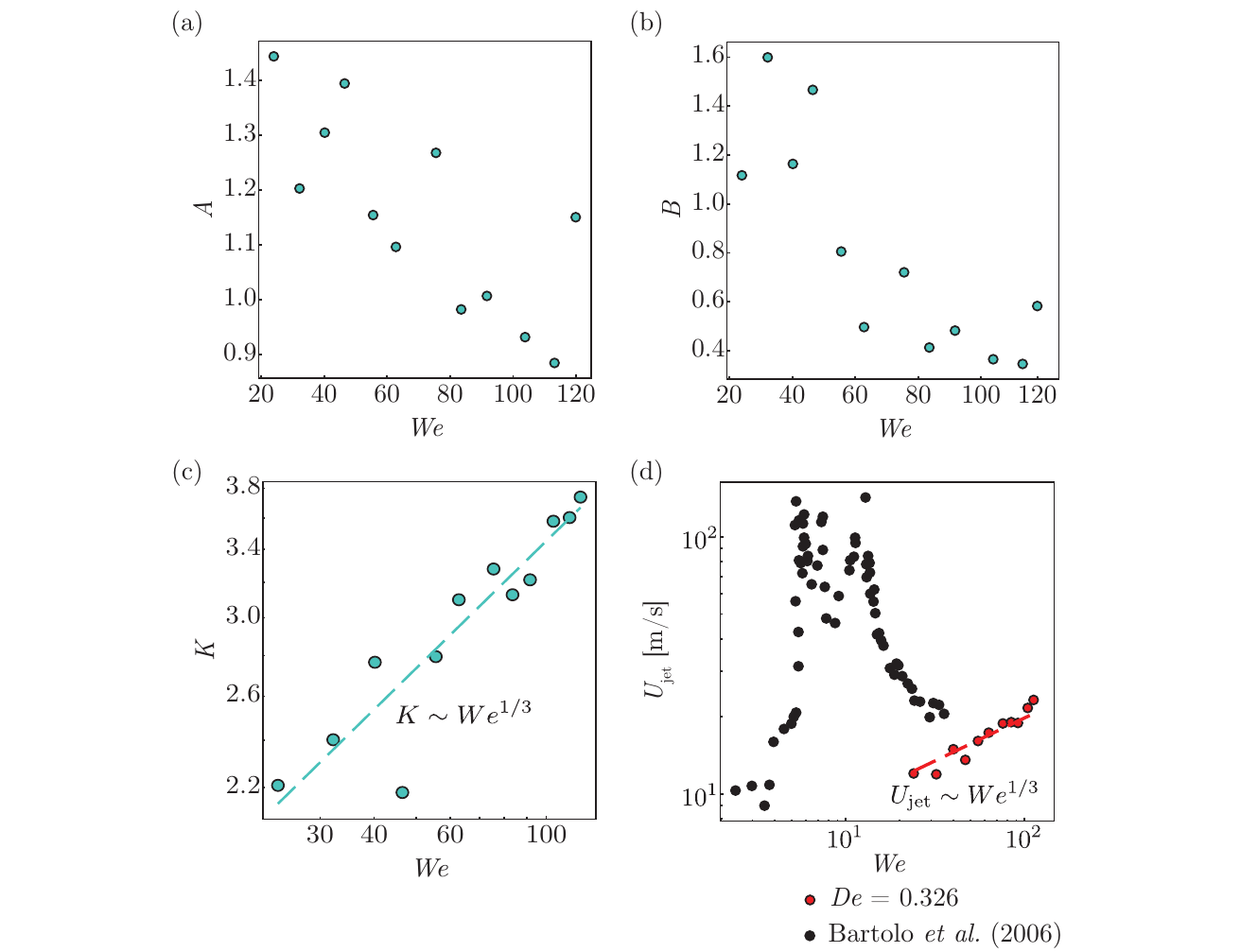}
\caption{Fitting coefficients (a) $A$ and (b) $B$ obtained from the fits~(\ref{eq:epsfit}) as functions of $W\!e$; (c) the prefactor $K$ (discrete datapoints) to the fits of $\dot{L}_{j}$~(\ref{eq:vel-fit}) is consistent with a $K \sim W\!e^{1/3}$ scaling; (d) variation of the initial jet velocity $U_{\text{jet}}$ as a function of $W\!e$. The dataset from \citet{bartolo-2006-prl} is also shown in panel d for comparison. For all the datapoints shown here: $De = 0.326$ and $Oh = 0.021$.}
\label{fig:app-fitting}
\end{figure}

\section{Shear viscosity measurements of the test liquids} \label{sec:app-rheo}

\begin{figure}
\centering
\includegraphics[width=\textwidth]{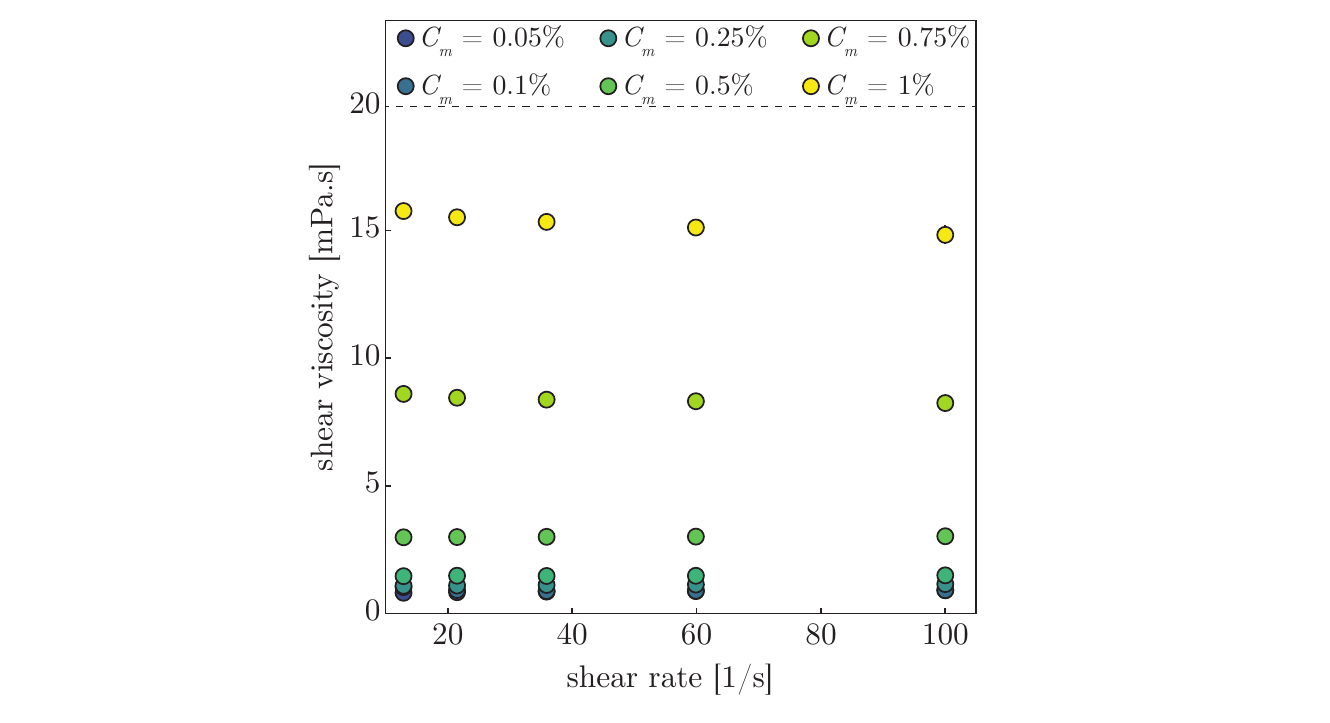}
\caption{Shear viscosity vs. shear rate measurements for all the test liquids in the present work, where $C_{m}$ indicates the polymer (PEO1M) concentration by mass.}
\label{fig:app-rheo}
\end{figure}

The variation of the shear viscosity of the polymeric liquids used in the present work as a function of the applied shear rate is shown in figure~\ref{fig:app-rheo}. The detailed measurement procedure has been described in \S~\ref{sec:exp-method}. As can be observed from figure~\ref{fig:app-rheo}, the shear viscosities of the test liquids are independent of the applied shear rate, thus indicating that our test liquids behave as Boger fluids \citep{james-2009-arfm}. 

\bibliographystyle{jfm}
\bibliography{ec-worthingtonjet-ref}

\end{document}